\documentclass[11pt]{article}
\usepackage{epsfig}
\begin{document}
\def\a{\alpha}\def\b{\beta}\def\g{\gamma}\def\d{\delta}\def\e{\epsilon }
\def\k{\kappa}\def\l{\lambda}\def\L{\Lambda}\def\s{\sigma}\def\S{\Sigma}
\def\Th{\Theta}\def\th{\theta}\def\om{\omega}\def\Om{\Omega}\def\G{\Gamma}
\def\y{\vartheta}\def\m{\mu}\def\n{\nu}
\def\ws{worldsheet}
\def\susy{supersymmetry}
\def\ts{target superspace}
\def\ks{$\k$--symmetry}
\newcommand{\plabel}{\label}
\renewcommand\baselinestretch{1.5}
\newcommand{\nn}{\nonumber\\}\newcommand{\p}[1]{(\ref{#1})}
\renewcommand{\thefootnote}{\fnsymbol{footnote}}
\thispagestyle{empty}
\begin{flushright}
hep--th/0403101
\end{flushright}

\bigskip
\thispagestyle{empty}

\vspace{1cm}
\begin{center}
{\Large\bf Diagrammar and metamorphosis of coset symmetries in
dimensionally reduced type IIB supergravity}

\vspace{2.5cm}
A. J. ~~Nurmagambetov
\footnote{Electronic address:  ajn@kipt.kharkov.ua}

\vspace{1.8cm}
{\small\it A.I. Akhiezer Institute for Theoretical Physics\\
NSC ``Kharkov Institute of Physics and Technology"\\
Kharkov, 61108, Ukraine}

\vspace{3.3cm} {\bf Abstract}
\end{center}
Studying the reduction of type IIB supergravity from ten to three
space-time dimensions we describe the metamorphosis of Dynkin
diagram for gravity line ``caterpillar" into a type IIB
supergravity ``dragonfly" that is triggered by inclusion of
scalars and antisymmetric tensor fields. The final diagram
corresponds to type IIB string theory $E_8$ global symmetry group
which is the subgroup of the conjectured $E_{11}$ hidden symmetry
group. Application of the results for getting the type IIA/IIB
T-duality rules and for searching for type IIB vacua solutions is
considered.

\vspace{3cm} {\it PACS: 04.65.+e, 04.50.+h, 11.25.Mj}

\renewcommand{\thefootnote}{\arabic{footnote}}
\setcounter{page}1

\newpage

One of the problems which has been actively studied last time is
to identify the underlying Superstring/M-theory symmetry group
\cite{westE11,westE8,ehtw}. Knowing the group is the essential
step in passing from higher energies and dimensions to
phenomenologically relevant vacua of M-theory and is the bypass
for bringing to light the hidden but essential ingredients of
higher dimensional theory that might help uncover its true
non-perturbative structure. The evidence on possible
group-theoretical ground of M-theory comes from studying the coset
symmetries in dimensionally reduced D=11 supergravity as the
M-theory low-energy limit. It was realized long ago \cite{cj} that
the global symmetry groups of D=11 supergravity toroidally
compactified down to four space-time dimensions fall into the
class of exceptional groups $E_n$ with $n\le 7$. Subsequent
reduction of D=11 supergravity to three \cite{MS} and to two
\cite{N} dimensional space-times (see also \cite{miz}) revealed
$E_8$ and $E_9$ global symmetry structure. The role of $E_{10}$ as
the hidden symmetry group of D=11 supergravity compactified onto
ten-dimensional torus was emphasized in \cite{julia82} and this
conjecture has been proved in \cite{miz}.

The important step in relating the global symmetry groups of
toroidally reduced D=11 supergravity to the true but hidden
symmetry group of the theory in eleven dimensions was done in
\cite{kns}. There was discovered the exceptional geometry of D=3
maximal supergravity that allowed to reformulate D=11 supergravity
in a $E_8$ invariant way. Together with previously obtained
results of \cite{dwn} and \cite{dtn} this observation gave the
strong evidence in favor of searching for the exceptional geometry
of M-theory based on a symmetry group which shall contain as a
subgroup the $E_n$ sequence of global low-dimensional symmetries
with $n\le 10$. Following previous experience it could be naively
expected to have, after ``compactifying" the time, a hidden
symmetry group whose algebra is of the rank eleven and includes
$E_9$ and $E_{10}$ as subalgebras. Since $E_{10}$ is the
hyperbolic Kac-Moody algebra \cite{GN} which contains $E_8$ and
its affine extension $E_9 \sim E^{+}_8$ as subalgebras and is
called the over-extension of $E_8$, i.e. $E_{10}\sim E^{++}_8$, an
M-theory hidden symmetry group should also be an extension of
hyperbolic Kac-Moody algebras. Such a generalization studied in
\cite{westE8}, \cite{gow}, \cite{ksw}, \cite{k} has been
christened the ``very-extension" of $E_8$ or $E_{11}\sim
E^{+++}_8$ (see also \cite{NO}, \cite{NF}).

The relevance of $E_{11}$ to the non-linear realization of
supergravities has been demonstrated for the bosonic subsectors of
higher dimensional maximal supergravities \cite{westE11},
\cite{sw}. A curious result is that the global hidden symmetry
group of type IIB theory turns out to be the same as for M\&A
theories related to each other by dimensional reduction. Since the
type IIB theory is not related to M-theory via straightforward
dimensional reduction on a circle the validity of $E_{11}$ as the
type IIB theory global symmetry group can be established in
studying the coset structure of dimensionally reduced type IIB
supergravity. Substantial fragments of this structure has been
found in \cite{cjlp}, \cite{cjlp0}, \cite{keur}, \cite{ksw},
\cite{sm}.

The aim of the present paper is to re-collect the fragments in a
systematic way during the application of ``anti-oxidation"
strategy of \cite{lw} and to emphasize the points that were
omitted from previous considerations. We are getting started with
ten-dimensional theory compactifying it to three space-time
dimensions. The special role of three-dimensional space-time in
establishing the hidden symmetries of higher-dimensional
supergravities has been emphasizing for a long time since there is
no gravity degrees of freedom in D=3 and dualising gauge fields we
get the theory whose dynamics is completely determined by scalar
degrees of freedom. Depending on the original higher-dimensional
sub-sectors of fields the scalars parameterize different $G/H$
coset spaces. Identifying the global group $G$ for different
subsectors of type IIB supergravity is our main task.

Let us begin our quest of the hidden symmetry group of type IIB
superstring theory with the following action describing dynamics
of the bosonic subsector of fields entering the type IIB
supergravity multiplet
$$
S=\int_{{\cal M}^{10}}\, [R\ast{\bf 1}+{1\over 2}d\phi_0\ast
d\phi_0 +{1\over 2}e^{2\phi_0}d\chi_0\ast d\chi_0
$$
$$
-{1\over 2}e^{-\phi_0}H^{(3)}\ast H^{(3)}-{1\over
2}e^{\phi_0}\tilde{H}^{(3)}\ast \tilde{H}^{(3)}+{1\over
4}\tilde{H}^{(5)}\ast \tilde{H}^{(5)}
$$
\begin{equation}\label{2b}
+{1\over 2}A^{(4)} dB^{(2)} dA^{(2)} +{\cal L}_{PST}].
\end{equation}
The first term of (\ref{2b}) corresponds to Einstein-Hilbert term,
$\int_{{\cal M}^{10}}\, \ast {\bf 1}\equiv \int\, d^{10}x
\sqrt{-g}$, $\phi_0$ and $\chi_0$ are the dilaton and axion
scalars. $H^{(3)}$ and $\tilde{H}^{(3)}$ are the field strengths
of NS and RR gauge fields $B^{(2)}$ and $A^{(2)}$
\begin{equation}\label{H3}
H^{(3)}=dB^{(2)},\qquad \tilde{H}^{(3)}=dA^{(2)}-H^{(3)}\chi_0.
\end{equation}
As well as the scalars they cast the doublet under $SL(2,R)$
global symmetry group of type IIB theory. $\tilde{H}^{(5)}$ is the
self-dual field strength of the $SL(2,R)$ singlet RR field
$A^{(4)}$
\begin{equation}\label{H5}
\tilde{H}^{(5)}=dA^{(4)}-H^{(3)} A^{(2)},\qquad
\tilde{H}^{(5)}=\ast \tilde{H}^{(5)}.
\end{equation}
The last term which is not important for the discussion in what
follows encodes the self-duality condition \cite{dls,dlt}. Since
we have made the choice of differential form notation the wedge
product between forms has to be assumed.

The first step in completing our task is to recover the following
structure of action
\begin{equation}\label{D3ac}
S=\int_{{\cal M}^3}\, \left[{1\over 2} d\vec {\phi}\ast d\vec
{\phi}+{1\over 2}\sum_{\vec{\alpha}}~e^{\vec{\alpha}\cdot
\vec{\phi}} d\chi_{\vec{\alpha}}\ast
d\chi_{\vec{\alpha}}\right]+\dots,
\end{equation}
which is obtained from (\ref{2b}) after performing dimensional
reduction on $T^7$. Here $\vec{\phi}=(\phi_0,\phi_1,\dots,\phi_7)$
is the dilaton vector comprised of the original dilaton $\phi_0$
and those appeared during dimensional reduction. The
$\vec{\alpha}$ are constant eight-vectors which label additional
(axionic) scalar fields $\chi_{\vec{\alpha}}$. The difference
between two types of scalars consists in sources of their
appearance due to dimensional reduction and different types of
interactions they possess \cite{popelec}. The axions come from
non-diagonal part of the Kaluza-Klein metric and from dualizing
the higher rank tensor fields. They possess only derivative
interactions. On the contrary the scalars associated with dilatons
comes from diagonal part of the metric and can have non-derivative
interactions as in (\ref{D3ac}). The form of the action
(\ref{D3ac}) is a signal that the scalars parameterize a $G/H$
coset space if of course one can identify the axion counting
vectors as positive roots of a group $G$. The global symmetry
group $G$ is uniquely defined by the Cartan matrix constructed out
of the simple roots \cite{popelec}.

To reach the action (\ref{D3ac}) we will use the same strategy as
in \cite{lw}. To this end one has to take into account the
standard rules of step by step toroidal reduction \cite{lpss}
$$
\int\, d^D x~eR \longrightarrow  \int\, d^{D-1} x~eR +\int_{{\cal
M}^{D-1}}[{1\over 2}d\phi_{1} \ast d\phi_1
$$
\begin{equation}\label{grdr}
+{1\over 2}e^{-2(D-2)\alpha_{D-1}\phi_1} F^{(2)}_1\ast F^{(2)}_1]
\end{equation}
with
\begin{equation}\label{alphDdf}
\sqrt{2 (D-2)(D-3)}\cdot \alpha_{D-1}=1,
\end{equation}
and
$$
\int_{{\cal M}^{D}}\, {1\over 2} F^{(n)}\ast F^{(n)}
\longrightarrow
$$
\begin{equation}\label{Fndr}
\int_{{\cal M}^{D-1}} [{1\over 2} e^{-2(n-1)\alpha_{D-1}\phi_1}
F^{(n)}_1 \ast F^{(n)}_1 + {1\over
2}e^{2(D-1-n)\alpha_{D-1}\phi_1} F^{(n-1)}_1 \ast F^{(n-1)}_1].
\end{equation}
The effect of transgression consisting in appearing new terms in
the reduced field strength $F^{(n)}_1=dA_1^{(n-1)}+\dots$ and of
having the Chern-Simons term in type IIB supergravity action are
denoted by the ellipsis in (\ref{D3ac}) and may be safely ignored
\cite{lw} since they have not an influence on the results in what
follows.

Let us get started with pure gravity case. Performing the
reduction in step by step manner one can recover six
seven-dimensional simple root vectors $\vec{\alpha}_k$ having the
following structure \cite{lw}
\begin{equation}\label{alphdf}
\vec{\alpha}_k= (0,\dots,0,-2(8-k)
\alpha_{9-k}
,2(6-k)
\alpha_{8-k}
,0,\dots,0)
\end{equation}
with $\alpha$'s from (\ref{alphDdf}), $k=0,\dots,5
$ zeros on the left and $(5-k)
$ zeros on the right. It is easy to verify that
\begin{equation}\label{aiak}
\vec{\alpha}_i\cdot \vec{\alpha}_k=
\cases{
4,& $i=k$ \cr -2,&$|i-k|=1$ \cr 0,& $|i-k|\ge 2$}
\end{equation}
All other roots coming from this subsector of type IIB
supergravity are not simple and can be expressed as a linear
combination of simple roots with non-negative coefficients.

One more simple root vector comes from dualising the Kaluza-Klein
vector field which appeared in the first step of reduction from
ten to three and having the following form
\begin{equation}\label{deldf}
\vec{\delta}=(16\cdot\alpha_{9},2\cdot\alpha_{8},2\cdot\alpha_{7},\dots,2\cdot\alpha_{3}).
\end{equation}
One can check that
\begin{equation}\label{delak}
\vec{\delta}\cdot \vec{\alpha}_k=
\cases{
0,&$k\ne 0$ \cr -2, &$k=0$, }
\qquad \vec{\delta}\cdot \vec{\delta}=4,
\end{equation}
and other roots that come from dualising the rest of the
Kaluza-Klein vectors are not simple. Denoting $\vec{\delta}$ as
$\vec{\alpha}_{(0)}$ one can construct the Cartan matrix
\begin{equation}\label{C0}
A_{ij}=2{\vec{\alpha}_i\cdot \vec{\alpha}_j \over \vec{\alpha}_i
\cdot \vec{\alpha}_i},\qquad i=(0),0,\dots,5
\end{equation}
that corresponds to the $A_{7}$ Dynkin diagram. This diagram is
that of $SL(8)$ group and is called the gravity line.

\vspace{1.2cm}
\begin{figure}[h]
\centering{\includegraphics
{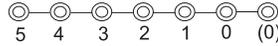}}
\caption{Gravity line ``caterpillar" of $A_{7}$.}
\end{figure}

Let us extend our analysis to include the dilaton-axion sector of
type IIB supergravity. Since we have $\phi_0$ from the beginning
we shall extend our simple root vectors (\ref{alphdf}),
(\ref{deldf}) with one additional column with zero on the left,
i.e.
$$
\vec{\alpha}_{(0)}=(0,16\alpha_{9},2\alpha_{8},2\alpha_{7},\dots,2\alpha_{3}),$$
etc. Hence we should deal with eight-dimensional simple root
vectors. And since we have axion $\chi_0$ from the beginning we
have also one additional eight-dimensional root vector
\begin{equation}\label{epsdf}
\vec{\epsilon}=(2,0,\dots,0).
\end{equation}
Clearly,
\begin{equation}\label{epsprod}
\vec{\epsilon}\cdot \vec{\epsilon}=4,\qquad \vec{\epsilon}\cdot
\vec{\alpha}_k=0,~\forall k.
\end{equation}
The Cartan matrix extended by new root corresponds to the
following Dynkin diagram

\vspace{1.3cm}
\begin{figure}[h]
\centerline{\includegraphics
{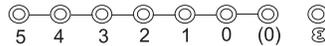}}
\caption{Beginning of the ``caterpillar's" metamorphosis.}
\end{figure}
that encodes the $SL(8)\otimes SL(2,R)$ group structure of this
sector of fields of type IIB supergravity. In the language of
Dynkin diagrams the first node on the right corresponds to the
$SL(2,R)$ group.

The next step is to extend our system of roots with inclusion of
$H^{(3)}$ and $\tilde{H}^{(3)}$ tensor fields. One more simple
root is coming from the reduction of the NS field kinetic term
\cite{lw}
\begin{equation}\label{betadf}
\vec{\beta}=(-1,12\cdot\alpha_{9},12\cdot\alpha_{8},0,\dots,0).
\end{equation}
It is easy to check that
\begin{equation}\label{betapr}
\vec{\beta}\cdot \vec{\alpha}_k=
\cases{
0,&$k\ne 1$ \cr -2,&$k=1$ \cr +2, &$k=(0)$, }
\quad \vec{\beta}\cdot \vec{\beta}=4, \quad \vec{\beta}\cdot
\vec{\epsilon}=-2.
\end{equation}
As such the root $\vec{\alpha}_{(0)}$ is not simple anymore since
the off-diagonal entries of the Cartan matrix are negative
integers or zero. It is merely a technical point to establish the
absence of other simple roots which could possibly come from the
dualisation of the NS 2-form gauge field and the rest of the
fields of type IIB multiplet and their dualisation. Therefore at
this stage of our study we arrive at the following diagram.

\vspace{1.5cm}
\begin{figure}[h]
\centerline{\includegraphics
{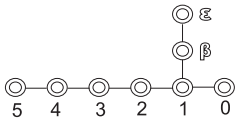}}
\caption{End of metamorphosis. ``Dragonfly" is fully fledged.}
\end{figure}

This diagram is topologically equivalent to the $E_8$ Dynkin
diagram and therefore the latter is the global symmetry group of
type IIB theory compactified to three space-time dimensions.

\vspace{0.6cm}
\begin{figure}[h]
\centerline{\includegraphics{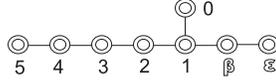}} \caption{Type IIB
$E_8$-like diagram.}
\end{figure}

Let us turn now to the applying the results obtained so far. We
will describe first the interpretation of T-duality rules in the
language of Dynkin diagrams (see also \cite{keur}, \cite{ksw}).
Doing the calculations outlined above one arrives at the following
diagram that encodes the coset structure of type IIA supergravity
reduced to three space-time dimensions. Here we have chosen a
slightly different notation to indicate the tensor fields from
which the simple roots came.

\vspace{0.2cm}
\begin{figure}[h]
\centerline{\includegraphics{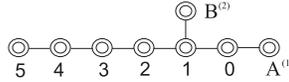}} \caption{Type IIA $E_8$
diagram.}
\end{figure}

Comparing Fig. 5 to Fig. 4 one can observe that two diagrams
coincide along the gravity line from the nodes 1 to 5. This is
just an indication of having the same gravity sub-sector for two
theories in D=9 space-time dimension. To have the same theories in
D=9 we have to identify the node corresponding to type IIA NS
2-form $B^{(2)}$ with that of type IIB gravity node, the type IIA
gravity node with the node of type IIB NS 2-form field and the one
of type IIA Kaluza-Klein vector field with that of type IIB axion.
This identification corresponds to seminal T-duality rules (cf.
e.g. \cite{ejl})
\begin{equation}\label{Trul}
i_z B^{(2)}_{IIA}\cong (i_Z g)_{IIB},\ \  (i_z g)_{IIA}\cong i_Z
B^{(2)}_{IIB},\ \  i_z A^{(1)}\cong \chi_0.
\end{equation}

Another important point in playing with Dynkin diagrams is the
possibility of identifying the relevant $AdS \times S$ vacuum
configurations \cite{sm}. Couple of years ago it was a
breakthrough in constructing the consistent non-linear
Kaluza-Klein ans\"atze for spherical dimensional reduction of
supergravities (see \cite{popelec} and \cite{cglp} for review).  A
systematic group-theoretical ground indicating the possibility of
such reductions is still lacking though it was formulated the
criterion of consistency of the reduction on $S^n$ based on the
possibility to enhance the global symmetry group after $T^n$
reduction due to ``conspiracy" of scalars.

Essential step in searching for the non-linear Kaluza-Klein
ans\"atze for spherical reductions is figuring out the possibility
of having the $AdS\times S$ vacuum configuration.
Recently it was proposed the method of examining such a
possibility based on considering the appropriate Kac-Moody
algebras \cite{sm}. In the context of type IIB supergravity the
evidence of $AdS_5 \times S^5$ vacuum configuration is based on
manipulations with $E_7$ diagram

\begin{figure}[h]
\centerline{\includegraphics{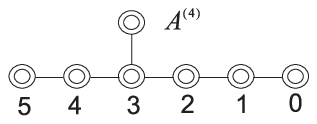}} \caption{Type IIB $E_7$
diagram.}
\end{figure}
extended with three additional nodes on the left, i.e. with
$E_7^{+++}$ diagram (cf. Fig. 7).

\vspace{0.6cm}
\begin{figure}[h]
\centerline{\includegraphics{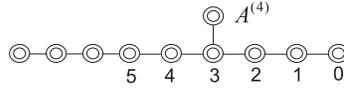}} \caption{$E^{+++}_7$
diagram.}
\end{figure}
The origin of the $E_7$ is easy to explain since the latter
corresponds to a ``larva" in the metamorphosis of Fig. 1 to Fig.
3. Such a diagram comes from the subsector of fields consisting of
gravity and self-dual 4-form gauge field. It is worth mentioning
that the root $\vec{\xi}$ coming from the reduction of
$\tilde{H}^{(5)}$ field strength and the one coming from its
dualisation obey the following relations
$$
\vec{\xi}\cdot \vec{\alpha}_k=
\cases{
0,&$k\ne 3$ \cr -2,&$k=3$ \cr +2, &$k=(0)$,}
\qquad \vec{\tilde{\xi}}\cdot \vec{\alpha}_k=
\cases{
0,&$k\ne 3$ \cr -2,&$k=3$ \cr +2, &$k=(0)$,}
$$
\begin{equation}\label{xi}
\vec{\xi}\cdot \vec{\xi}=4,\qquad \vec{\tilde{\xi}}\cdot
\vec{\tilde{\xi}}=4,\qquad \vec{\xi}\cdot \vec{\tilde{\xi}}=4.
\end{equation}
Hence it is a matter of taste which one is selected to be the
simple root that is the remnant of the $\tilde{H}^{(5)}$
self-duality. As soon as the choice was made another root is no
longer simple.

Skipping the details of manipulations with $E^{+++}_7$ diagram
that leads to the $AdS_5\times S^5$ configuration (we refer the
reader to the original paper \cite{sm}), it is worth mentioning
that the subset of fields leading to the $E^{+++}_7$ is precisely
the one for which the existence of the non-linear ansatz for the
$S^5$ dimensional reduction \cite{popeS5} was proved! Another
example of having the non-linear ans\"atze for $AdS_{D-3}\times
S^3$ and $AdS_3\times S^{D-3}$ configurations was established for
the bosonic string theory \cite{popeS3} which includes graviton,
dilaton and 2-rank gauge field in the massless sector. This
subsector enters the type IIB supergravity and corresponds to the
$D^{+++}_8$ Kac-Moody group. One can verify following the approach
of \cite{sm} that such vacuum solutions are indeed the case.

To summarize, we have traced the metamorphosis of Dynkin diagrams
representing the symmetries of different sub-sectors of
dimensionally reduced type IIB supergravity. This provides the
link to the results obtained in the framework of non-linear
realization of type IIB supergravity \cite{sw} and of searching
for the M-theory hidden symmetry group
\cite{westE11,westE8,ehtw,ksw,k} as well as to the results
obtained by use of the oxidising technique \cite{cjlp0,keur}. The
graphical representation of the coset symmetries in dimensionally
reduced supergravities encodes a lot of information on the matter
field content of a theory, the relevant low-dimensional vacua and
dualities between different supergravities and is therefore the
very useful tool in studying the hidden symmetry structure of
Superstring/M-theory.

{\bf Acknowledgements.} We are very grateful to Igor Bandos and
Dmitri Sorokin for pleasant discussions and constant
encouragement. This work is supported in part by the Grant N
F7/336-2001 of the Ukrainian SFFR and by the INTAS Research
Project N 2000-254.

\end{document}